\documentclass[%
twocolumn,amsmath,amssymb
reprint,
superscriptaddress,
amsmath,amssymb,
prl,
]{revtex4}
\usepackage{xcolor}
\usepackage{graphicx}% Include figure files
\usepackage{dcolumn}% Align table columns on decimal point
\usepackage{bm}% bold math
\usepackage{float}
\usepackage{threeparttable}
\usepackage{upquote}  % for comments
\usepackage{multirow}
\def\beq{\begin{equation}}
\def\eeq{\end{equation}}
\def\bea{\begin{eqnarray}}
\def\eea{\end{eqnarray}}
\def\brcl{\begin{array}{rcl}}
\def\bccl{\begin{array}{ccl}}
\def\blcl{\begin{array}{lcl}}
\def\err{\end{array}}

\begin{document}

\preprint{APS/123-QED}

\title{Exploring Chemical Compound Space with Quantum-Based Machine Learning}

\author{O. Anatole von Lilienfeld}
\email{anatole.vonlilienfeld@unibas.ch}
\affiliation{Institute of Physical Chemistry and National Center for Computational Design and Discovery of Novel Materials (MARVEL), Department of Chemistry, University of Basel, 4056 Basel, Switzerland}
\author{Klaus-Robert M\"uller}
\email{klaus-robert.mueller@tu-berlin.de}
\affiliation{Machine Learning Group, Technische Universität Berlin, 10587 Berlin, Germany}
\affiliation{Korea University, Department of Brain and Cognitive Engineering, Seoul, Korea }
\affiliation{Max Planck Institute for Informatics, Saarbr\"ucken, Germany}
\author{Alexandre Tkatchenko}
\email{alexandre.tkatchenko@uni.lu}
\affiliation{Physics and Materials Science Research Unit, University of Luxembourg, L-1511 Luxembourg, Luxembourg}

\date{\today}% It is always \today, today,

\begin{abstract}
Rational design of compounds with specific properties requires conceptual understanding and fast evaluation of molecular properties throughout chemical compound space (CCS) -- the huge set of all potentially stable molecules. 
Recent advances in combining quantum mechanical (QM) calculations with machine learning (ML) provide powerful tools for exploring wide swaths of CCS. 
We present our perspective on this exciting and quickly developing field by discussing key advances in the development and applications of QM-based ML methods to diverse compounds and properties and outlining the challenges ahead.
We argue that significant progress in the exploration and understanding of CCS can be made through a systematic combination of rigorous physical theories, comprehensive synthetic datasets of microscopic and macroscopic properties, and modern ML methods that account for physical and chemical knowledge.      
\end{abstract}

\pacs{Valid PACS appear here}% PACS, the Physics and Astronomy
\maketitle

%\tableofcontents

\section{Introduction}
Due to an unfathomably large number of possible molecules and materials~\cite{ChemicalSpace,mullard2017drug}, and the combinatorially many ways for them to undergo chemical transformations, our understanding of chemistry requires a first-principles approach with proper roots in quantum mechanics (QM) and statistical mechanics (SM). QM gives us the ability to calculate accurate microscopic properties (energies, atomic forces, electronic energy levels, electrostatic multipoles, polarizabilities) for fixed molecular geometries, while SM allows us to sample QM energy surfaces in a given statistical ensemble and calculate macroscopic properties. Accurate QM and SM simulations are computationally demanding, even for a single molecule or material, hence more efficient approaches are urgently needed to address the length- and time-scale dilemma in molecular simulations with sufficient accuracy in order to obtain insights into the evolution of properties throughout CCS. Such efficient methods may eventually enable the long-held dream of \emph{in silico} chemical and materials design.

\begin{figure}
\centering
\includegraphics[width=8.5cm]{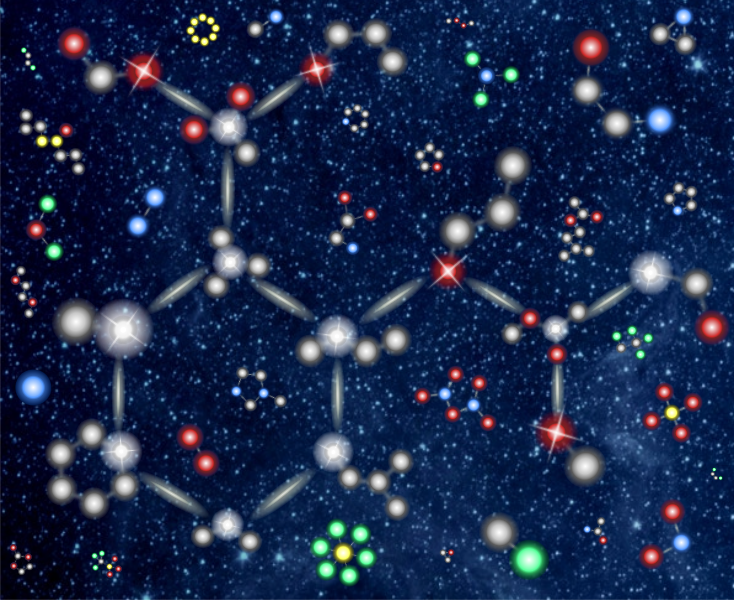}
\caption{
{\color{black} The structure of aspirin emerges, in analogy to a stellar constellation, as a superposition of certain atoms (highlighted as stars) in their mutual molecular environments, and amid a cloud of unrelated molecules. 
Nowadays, most QML approaches decompose predicted properties into atomic contributions. For example, computing quantum properties, such as the atomization energy, as a series expansion in select fragments called ``amons'' has by now enabled learning properties for chemically diverse molecules. 
%For each property, the regression coefficients of the corresponding amon contributions differ. 
Amon stands for an \underline{A}tom in a \underline{M}olecule, and the suffix \underline{on} indicates that it can be used as a building-block dictionary entry which is selected whenever relevant in the ensemble of constituting fragments of any larger query molecule~\cite{Amons}.
}  
Gray, red, blue, green, and yellow encode carbon, oxygen, nitrogen, fluorine, and sulfur, respectively.
Hydrogen atoms, as well as single and double bonds have been omitted for clarity. 
%Illustration of chemical space as a molecular analogy to a stellar constellation. 
%The molecular graph of aspirin emerges from a combination of some of its constituting atomic fragments (so-called ``amons'')~\cite{Amons} (highlighted as stars), amid a cloud of unrelated molecules.
%Colors encode elemental identity.
%J.~Heweliusz superimposed with molecular sub-graphs of aspirin molecule.
}
\label{fig:Constellation}
\end{figure}

In practice, we refer to CCS as the set of all feasible metastable atomic configurations resulting from solving Schr\"odinger's equation. Non-equilibrium molecular configurations provide smooth interpolations between points in this high-dimensional chemical space. 
CCS is large but finite: While accounting for all possible conformations of all stereoisomers of all possible constitutions of all possible compositions, it also encodes repeating patterns, abundant signatures, and low-dimensional building-blocks.  
Fig.~\ref{fig:Constellation} illustrates the pervasiveness of the underpinning constituting patterns when
charting CCS by drawing an analogy to stellar constellations.
Certain star constellations carry names, and so do certain molecules. 
More importantly, stellar patterns have been useful for orientation and navigation. 
{\color{black} Similarly, property patterns throughout chemical space can be combined in ``constellations'', from which properties of new molecules of interest can be calculated using linear or nonlinear combination of properties of known molecules or molecular fragments.}
While relationships for stars and planets are rather well understood, a rigorous understanding of CCS in terms of molecular components and to a similar degree has not yet been achieved but would be of utmost usefulness for rational compound design. In this perspective, we argue that the recently developed machine learning (ML) approaches will significantly aid in achieving a deeper understanding of CCS. We give several examples that illustrate the substantial progress achieved in this field, and outline the many remaining challenges yet to be addressed. 

Fundamentally speaking, QM describes the electronic structure of any material compound, thereby determining the behavior of matter at large, and dictating all the mutual relationships between observable microscopic properties~\cite{anatole-ijqc2013}. 
In light of such a large domain of applicability it is not a surprise that fundamental contributions to density-functional theory -- one of the more efficient formulations of QM -- are among the top cited papers of all times~\cite{van2014top}.
The unbiased study of chemical space for the purpose of exploration as well as exploitation (computational compound design) imposes a severe need for sampling algorithms with maximal efficiency. While QM based design of materials has already been successfully applied to some specific materials design challenges~\cite{ZungerNature1999,NorskovPRL2002,Curtarolo2013}, it imposes a prohibitive computational cost in general. Consequently, the improvement in efficiency and robustness of electronic-structure calculations will play an increasingly important role in current and future materials design efforts~\cite{ComputationalMaterialsDesign_MRS2006,MGI2011,marzari2016materials,alberi20182019}.

The targeted exploration of CCS aiming to obtain compounds with desired properties is a longstanding endeavor. Many efforts in cheminformatics or materials informatics have relied on statistics and ML to search CCS for relevant pharmaceutical properties such as receptor binding, toxicity~\cite{ADMET_Oprea,butina2002predicting} or materials stability~\cite{rajan2005materials,MachineLearningHautierCeder2010,ward2017atomistic}. 
While often useful and computationally efficient, the main drawback of these approaches is that they are not transferable to molecules and properties outside of their domain of applicability, which results from a lack of underlying principles of physics, as also pointed out in Ref.~\cite{SchneiderReview2010}.

Here instead we focus on recent fundamental ML developments aimed at a quantum-based understanding of CCS. The key idea is that any observable property for any system can be obtained from solving the relevant quantum-mechanical equations. 
A more comprehensive QM-based understanding of CCS is now foreseeable because of (i) maturity, efficiency \& reproducibility of electronic structure methods {\em and} codes, (ii) fast-paced developments in high-performance computing hardware, and (iii) conceptual adaptations of statistical mechanics and statistical learning techniques. 
Regarding (i), we note the many substantial theory and method developments including density-functional theory (DFT) and post-Hartree-Fock (HF) wavefunction methods, as well as the significant efforts made to ensure reproducibility across different electronic-structure codes~\cite{lejaeghere2016reproducibility}.
Regarding (ii), the scientific codes in the electronic-structure community have matured to such an extent that a considerable fraction of the world's top high-performance computing centers busy themselves with QM calculations. 
Regarding (iii), continuous advances in statistical learning have nowadays enabled performing intelligent data analysis on both small and large data sets, and to extract valuable quantitative insights in a systematic manner. We note in passing that machine learning has already enabled many applications in other fields~\cite{lecun2015deep,schmidhuber2015deep}, including medical diagnostics~\cite{capper2018dna,klauschen2018scoring,jurmeister2019machine}, particle physics~\cite{baldi2014searching}, bioinformatics~\cite{lengauer2007bioinformatics}, brain-computer interfaces~\cite{blankertz2008optimizing}, social media analysis~\cite{perozzi2014deepwalk}, robotics~\cite{thrun2005probabilistic} and team, social, or board games~\cite{Moneyball,IBMWatson,silver2016mastering}.

Many recent publications, including special journal issues and reviews of quantum-based ML approaches~\cite{rupp2018guest,MLvirtualissue2018jpc,QMLessayAnatole,batista2019designMLalchemy}
have highlighted the fact that combining quantum calculations with ML can lead to 
considerable leaps in exploring and understanding chemical and materials spaces. 
In addition to learning quantum-mechanical observables (integrating out electronic degrees of freedom), 
evidence has been presented that it is equally possible to build ML models of SM ensemble averages (integrating out the atomistic
configurational degrees of freedom), such as free energy,
entropy, or kinetic pathways~\cite{DeltaPaper2015,mardt2018vampnets}.

To distinguish this emerging field of physics-based machine learning from 
preceding efforts in chemo-~, bio-, and materials-informatics we will refer to combinations of QM and SM approaches 
with machine learning as QML models in the following.
QML refers to the idea of applying modern statistical learning theory to predict electronic and atomistic properties and processes in molecules and materials. We also remark that the goals and reaches of QML models should not be confused with quantum ML algorithms executed on quantum computers. As such, QML models aim to provide a feedback mechanism between QM/SM and (statistical) machine learning. Given sufficient reference data obtained from QM and SM simulations, queries of properly trained ML models can yield accurate properties within milliseconds---as opposed to the many CPU hours or days necessary to solve the corresponding quantum and statistical mechanics problems for representative compounds. Since QML models interpolate rigorously in complex nonlinear spaces with controlled predictive accuracy, the door has now opened for more extensive analysis and study of these interpolated spaces, previously impossible due to the prohibitive computational cost of direct QM and SM simulations.    

Given the substantial progress in the QML field discussed in this perspective article, we argue that significant progress in the exploration and understanding of CCS can be made through systematic combination of rigorous physical theories, comprehensive datasets of QM and SM properties, and sophisticated ML methods that incorporate physical and chemical knowledge. The authors have witnessed the quick development of the QML field from the perspective of electronic structure calculations and hence the focus in this perspective will be on combining QM and ML with the goal of enhanced exploration of CCS.
{\color{black} Efforts to use ML to capture SM properties in analogous ways are subject of active current research~\cite{noe2018machine,FrankBoltzmann}.}

\section{Box1: Explanation of QML terms}
Here we give compact explanation for various key-words discussed in this perspective: 
\begin{itemize}
  \item Chemical Compound Space (CCS): The set of all feasible metastable atomic configurations resulting from solving the Schr\"odinger equation for the corresponding system of interacting electrons and nuclei.
  \item Machine Learning (ML): Methods based on statistical learning theory for obtaining numerical models from data samples that generalize well on unseen data. Generally, ML models improve with the availability of more data, hence the models are said to ``learn from data''. ML models are inductive, i.e. they are typically not based on any underlying physical model, but can, in principle, reconstruct physical models from the provided data. In the context of exploring CCS, data is less abundant than e.g. in typical ML applications like computer vision. It becomes therefore important to most efficiently make use of available data by combining prior knowledge about physics and chemistry with powerful ML models, as we will argue throughout this perspective. 
  
   \item Representation: The model that encodes the structure of and relations between atoms. It is crucial for quantifying similarities. For molecules, the representation needs to be unique and invariant to atom indexing as well as to molecular translations and rotations in space (see Fig.~\ref{fig:Rep}). 
  
  \item Supervised, unsupervised, and semi-supervised learning: Machine learning with labels is called supervised learning (SL). Examples of SL are classification or regression where for every sample its class label or regression value is given in the training data. Unsupervised learning (USL) on the contrary has no label information; clustering and dimensionality reduction are typical USL problems. Semi-supervised learning assumes that most samples have no label, and only for very few samples labels are provided for training. 
  
  \item Parametric and nonparametric models: Parametric models assume a finite set of model parameters that need to be estimated (e.g.~a mean of a Gaussian), whereas nonparametric models do not make this assumption. Popular nonparametric models such as Gaussian processes can be viewed as having infinitely many parameters. 
  
   \item Regression: In regression the relationship between an input representation and continuous output variables is estimated. The most common simple regression analysis is linear regression where a linear function (in one dimension a line, in $N$ dimensions a hyperplane) is fitted to the data according to some loss function such as mean squared error. A widely used classical model for non-linear regression is Kernel-Ridge Regression (KRR) that generalizes well to unseen data with limited scalability in larger data sets.
 
  \item (Deep) Neural Networks (DNNs): Widely used and flexible non-linear regression models based on neural networks. Deep networks refer to structured architectures that have a large number of hidden layers offering large flexibility and rich multiscale representations. Due to their scalability they are ideally suited to extract complex non-linear relations from large data sets. 
  \item Cross validation: A common ML procedure used for ensuring generalization to unseen data and avoiding overfitting. 
  \item Learning curves: Measure the performance of ML models upon increasing the number of data samples used for training the model.
  \item Density-Functional Theory (DFT): The workhorse method for electronic-structure calculations on molecules and materials. While DFT is in principle an exact theory, in practice approximations are made for electronic quantum (exchange and correlation) effects. DFT implementations often provide a good compromise between accuracy and efficiency. Most current QML datasets for molecules and materials are based on DFT calculations. 
  
  \item Hartree-Fock (HF): Fundamental electronic-structure method used as a starting point for essentially all practical calculations of quantum correlation energy in molecular systems. HF provides an exact treatment of electronic exchange effects due to Pauli repulsion.
  \item Coupled cluster (CC) methods: A set of methods to obtain an systematically improve the calculated estimate of electronic correlation energy based on the HF wavefunction. This is achieved by increasing the level of modeled electronic excitations: single excitations, double excitations, and so on. However, higher level treatment requires orders of magnitude more computational resources. 
  \item CCSD(T): The so-called "gold standard" method of computational quantum chemistry, including single, double, and perturbative triple excitations (to fourth order in perturbation theory). Within converged basis-sets, CCSD(T) typically yields so-called "chemical accuracy" of $\sim$1 kcal/mol in atomization energies of molecules with single reference character.
\end{itemize}

\begin{figure}
\centering
\includegraphics[width=8.5cm]{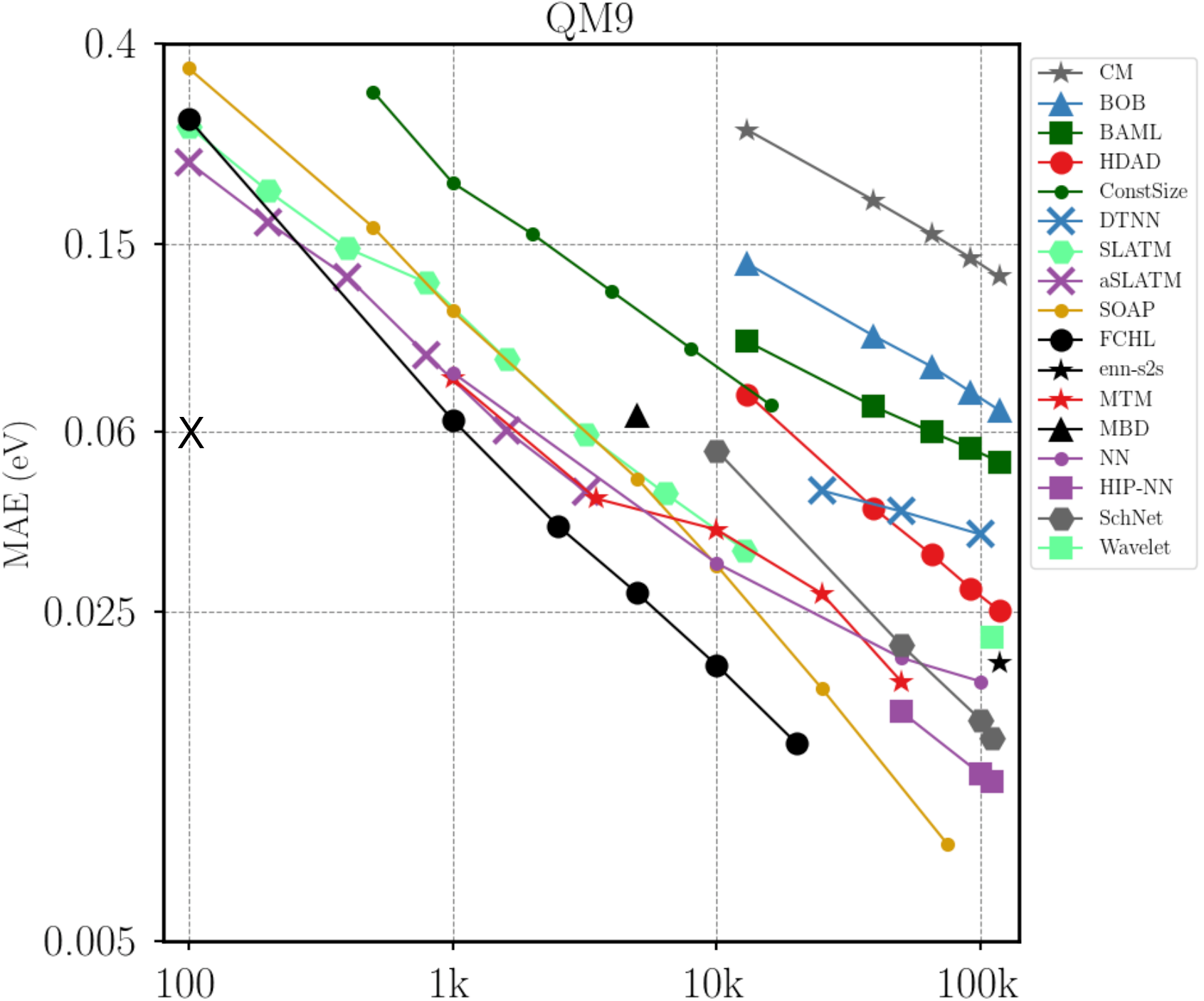}
\caption{
Learning curves illustrate last years' progress of QML models of atomization energies of molecules.
This plot shows mean absolute error (MAE) in eV on atomization energies of small molecules in the QM9 dataset~\cite{QM9}.
Shown QML models differ solely by representation and model architecture, and correspond to CM (2012)~\cite{CM}, BOB (2015)~\cite{BOB}, BAML (2016)~\cite{BAML}, HDAD (2017)~\cite{HDAD}, constant-size-descriptors (2018)~\cite{collins2018constant}, DTNN (2017)~\cite{DTNN2017}, 
(a)SLATM~\cite{Amons} (2017), SOAP (2017)~\cite{CeriottiScienceUnified2017}, FCHL (2018)~\cite{FCHL}, enn (2017)~\cite{NeuralMessagePassing}, MTM (2018)~\cite{gubaev2018machine}, MBD (2018)~\cite{pronobis2018many}, NN (2018)~\cite{unke2018reactive}, HIP-NN (2018)~\cite{nebgen2018transferable}, 
SchNet (2018)~\cite{SchNet}, Wavelets (2018)~\cite{eickenberg2018solid}.
The black X indicates the so-called ``QM9 challenge'' (developing a QML model capable of achieving 1 kcal/mol (0.043 eV) accuracy on the QM9 dataset using only information of 100 molecules for training, see {\tt https://tinyurl.com/y2e589wj} for more details) formulated at the UCLA-IPAM workshop in 2018. To date, this challenge has not been met.
}
\label{fig:QML}
\end{figure}

\begin{figure*}
\centering
\includegraphics[width=17cm]{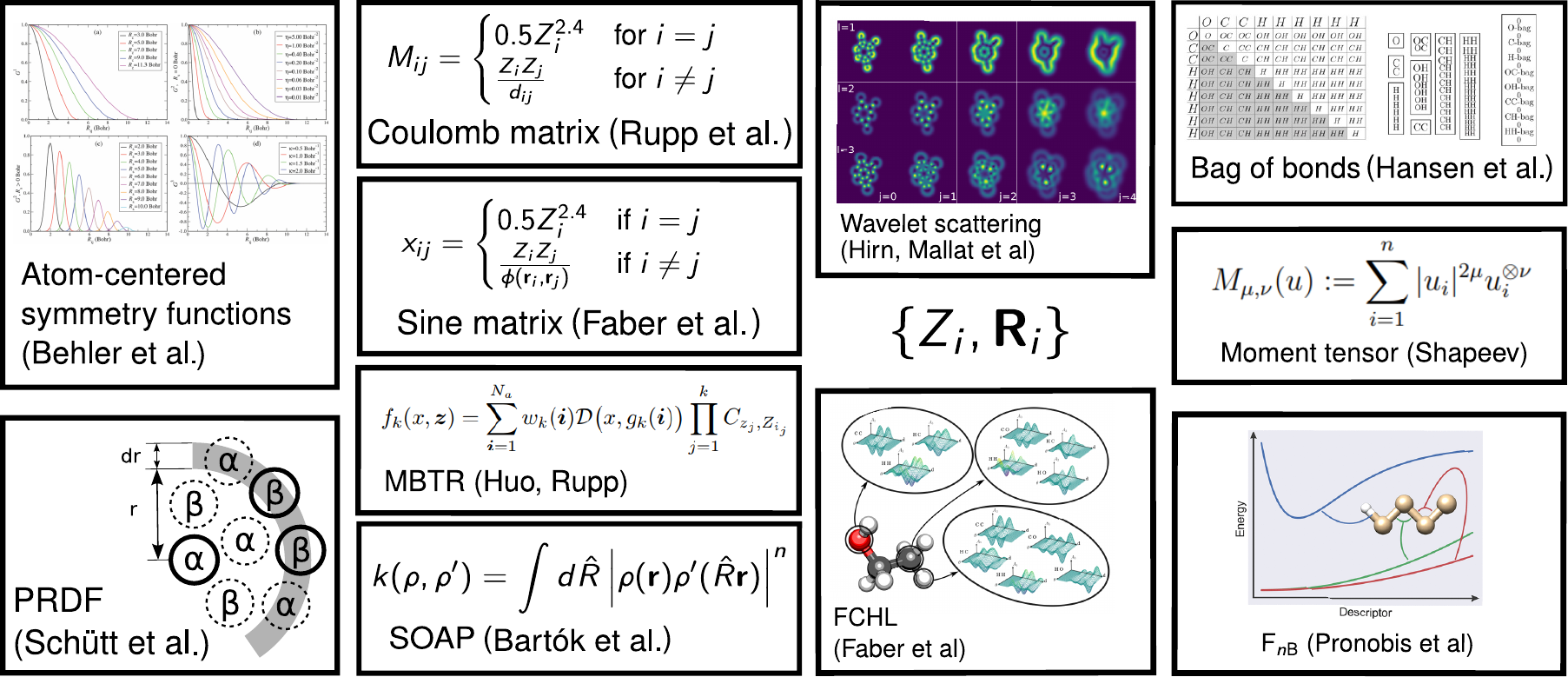}
\caption{
Illustration of several widely used representations for kernel-based ML of molecules and materials:
Atom-centered symmetry functions (Behler \emph{et al.})~\cite{Behler_NNreview2015},
Coulomb matrix (Rupp \emph{et al.})~\cite{CM}, Bag-of-Bonds (Hansen \emph{et al.})~\cite{BOB}, SOAP (Bartók and Csanyi \emph{et al.})~\cite{CeriottiScienceUnified2017}, PRDF (Schütt \emph{et al.})~\cite{schutt2014represent},
FCHL (Faber \emph{et al.})~\cite{FCHL}, MBD (Pronobis \emph{et al.})~\cite{pronobis2018many}, MBTR (Huo and Rupp)~\cite{RuppsMBTR2017}, 
Wavelets (Hirn and Mallat \emph{et al.})~\cite{eickenberg2018solid}, Moment tensor (Shapeev)~\cite{shapeev-tensor}, 
Sine matrix (Faber \emph{et al.})~\cite{MLcrystals_Felix2015}.
}
\label{fig:Rep}
\end{figure*}

\section{Goals and advances of QML}
The overarching goal of QML is to develop reliable models with accuracy of high-level electronic structure calculations. Depending on the application, the reference data can be obtained from high-level quantum chemistry, i.e. CCSD(T), or from DFT calculations. While much work remains to be done to reach the ``dream'' of exact QML models, many key advances have been recently achieved that we will review in this section and connect to important remaining challenges where we deem that urgent progress is needed. 

The first salient point is that all QML advances hinge on the availability of trustworthy QM data. This data needs to cover a certain important domain, for example the CCS of organic drug-like compounds, as explored by
Reymond and co-workers through their GDB list of SMILES strings~\cite{ReymondChemicalUniverse,ReymondChemicalUniverse2,ReymondChemicalUniverse3, GDB17}. 
Subsequent QM calculations on these molecular graphs led to the
publication of equilibrium structures and properties of many thousands of small molecules (QM7 and QM9 datasets~\cite{Montavon2013,QM9}), 
their molecular dynamics trajectories (MD17 dataset~\cite{chmiela2017machine}), or non-equilibrium molecular structures (ANI-1 dataset~\cite{smith2017ani}). 
On the side of inorganic materials, one can calculate equilibrium structures of solids (AFLOW, OQMD, and Materials Project datasets~\cite{MaterialsProject,OQMD,Elpasolite_2016}), 
or generate equilibrium and non-equilibrium MD data for a single element (for example silicon~\cite{bartok2018machine}). 
The eventual goal of QML is to develop a universal efficient model for all of CCS capable of accurately describing molecular and materials data on equal footing, and possibly leading to new insights on its underlying regularity and chemical relationships. Reorganizing the periodic table represents a first and important stepping stone in that direction~\cite{Elpasolite_2016,willatt2018feature}.
Initially, different models have been developed focusing either on molecular or materials data, while more recent flexible developments can already be applied to both molecules and solids~\cite{FCHL,SchNet,CeriottiScienceUnified2017}.    

%ANATOLE: 
While CCS is also commonly explored within cheminformatics based approaches,
QML differs in the sense that it rigorously adheres to its roots in fundamental physics, i.e. that it is consistent with the laws of QM and SM. 
Probably the first QML applications employing ML techniques for non-linear interpolation of QM data were aiming to construct reliable system-specific interatomic potentials, or potential-energy surfaces, going beyond conventional force fields~\cite{SumpterNoidNeuralNetworks1992,Rabitz1996,Neuralnetworks_Scheffler2004,Neuralnetworks_BehlerParrinello2007,bpkc2010,BehlerPerspective2016}.

Then in 2011, QML was extended to develop a transferable model for describing QM properties, trained and applicable throughout CCS as demonstrated for the QM7 set of organic molecules~\cite{CM}, highlighting the potential of QML for efficient and accurate exploration of CCS. This idea was rapidly demonstrated to be applicable to many electronic properties using neural networks as well as kernel ridge regression~\cite{Montavon2013,AssessmentMLJCTC2013,SingleKernel2015}, or to search for polymers with useful properties~\cite{ML4Polymers_Rampi2013}, explore chemical properties of crystalline solids~\cite{schutt2014represent,ML4Crystals_Wolverton2014,Elpasolite_2016,WolvertonMLcrystals2017,GrossmannCNN2018}, and design promising materials for a variety of technological applications~\cite{Alan_OLED2015,jorgensen2018exploration}.

A crucial aspect that determines the reliability and applicability of any QML model is its generalization accuracy assessed on the calculated QM properties of a sufficiently large out-of-sample (hold-out) dataset. It is remarkable how quickly the generalization accuracy of QML models has improved during the past few years. As shown in Figure~\ref{fig:QML} on the example of QM9 dataset, the QML errors have decreased by 40-fold -- from 8 kcal/mol in 2012 to 0.2 kcal/mol in 2018, using exactly the same dataset for training and validation. This noticeable increase in accuracy mainly stems from incorporation of physical prior knowledge into the QML models, such as proper description of permutational symmetries of atoms in a molecule~\cite{CeriottiScienceUnified2017,FCHL,chmiela2017machine,chmiela2019sgdml},
as well as explicit inclusion of physically-motivated pairwise and many-body terms for interatomic interactions into QML descriptors~\cite{BAML,pronobis2018many}. An illustration of several widely used representations of molecules in the context of kernel-based ML is shown in Fig.~\ref{fig:Rep}. A noticeable conclusion from Figure~\ref{fig:QML} is that the most advanced existing QML models are extremely data efficient, achieving chemical accuracy of 1 kcal/mol for QM9 dataset of 134'000 molecules using only 1000 molecules (0.7\%) for training. This result hints on the potential sparsity of CCS, implying a low complexity and dimensionality~\cite{Braun2008} of the property prediction problem throughout CCS. In other words, this evidence suggests that (unknown) properties of query molecules can be predicted as non-linear combination of (known) properties of only few other molecules, which are not necessarily chemically similar. In order to further the search for improved QML models, the authors are among a group of professors who each agreed to award US\$100 to the scientist(s) who devise(s) a QML model which meets the IPAM ``QM9 challenge'' [{\tt https://tinyurl.com/y2e589wj}] (See Fig.~\ref{fig:Rep}).  

A critical component of every QML model lies in the representation (sometimes also referred to as ``descriptor'') of an atomic system composed of nuclei and electrons. 
Uniqueness of the representation is a necessary condition for QML models to converge~\cite{FourierDescriptor}, and an increase in similarity with respect to the potential-energy surface, i.e.~the target function, lowers the offset in learning curves~\cite{BAML}.
In order to afford efficient learning, representations should also capture well-known translational, rotational, and permutational invariances of atomistic systems, but also can be significantly improved by explicitly including other physical priors, such as temporal and spatial symmetries of interatomic interactions~\cite{chmiela2017machine}, or differential relationships of response properties~\cite{christensen2018operators}. 
An additional requirement is that atomistic representations should be as computationally efficient as possible in order to reap the benefits of QML's computational efficiency over the numerical solvers employed within conventional QM calculations. 
While both unified~\cite{RuppsMBTR2017} as well as specific~\cite{CM,BartokGabor_Descriptors2013,BOB,schutt2014represent,MLcrystals_Felix2015,pronobis2018many} atomistic
representations have been proposed in the literature, there is still an ongoing debate and empirical assessment of the advantages and limitations of different representations for particular application domains. 
While kernel-based machine learning models require explicit formulation for the representation~\cite{CM,AssessmentMLJCTC2013,BartokGabor_Descriptors2013} using one scale per kernel, deep neural networks such as SchNet or DTNN~\cite{SchNet,DTNN2017,unke2018reactive} are able to construct an implicit multiscale representation as an outcome of a scalable learning process. 

In QM calculations using quantum chemistry or DFT, different properties of an atomistic system (electronic energy, atomic forces, multipole moments, polarizability, electronic energy levels) can be evaluated as QM operators acting on the electronic wavefunction. The situation is more intricate in the case of QML models. Initially, separate QML models were used to describe different properties, i.e. one model for total energy, and another one for the polarizability. A first success of neural networks applied to QM7 dataset was to demonstrate the transferability of a single NN to predict multiple electronic properties at the same time~\cite{Montavon2013}. In other words, the NN had learned an invariant representation, universal enough to capture several properties simultaneously. Since then, kernel methods and deep neural networks have also been extended to reflect multi-property prediction~\cite{SingleKernel2015,IsayevSciAdv2019}. However, beyond that it would be very advantageous to view QML models as a coarse-grained surrogate for electronic interactions in an atomistic system, akin to a downfolded version of the wavefunction. In such a model, electronic properties in QML could be calculated similarly to explicit QM calculations, namely as operators on manifolds. This idea of using response operators for ML in chemical space has been recently introduced by Christensen \textit{et al.}~\cite{christensen2018operators}.
{\color{black} Various QML models of electron densities and wavefunctions have by now also been introduced~\cite{ML4Kieron2012,QMB_ANN2017,brockherde2017bypassing,schnorb,fabrizio2019electron,hermann2019deep,pfau2019abinitio}
}

In general, the chemical space of molecules and materials can be expanded into compositional (chemical elements forming bonds) and configurational (same bonds, but different structures) degrees of freedom. Up to now, we have mainly discussed models that explore compositional degrees of freedom in CCS. On the other hand, configurational degrees of freedom are crucial to understand dynamics of molecules under given external conditions. QML models have also been successfully employed for the construction of advanced interatomic potentials for MD applications that aim beyond the realm of classical, system-dependent force fields~\cite{BartokGabor_Descriptors2013,Behler_NNreview2015,SchNet,chmiela2017machine,CeriottiScienceUnified2017,shapeev-tensor}. As indicated earlier, the construction of reliable ML force fields from QM data was one of the first examples of successful applications of QML~\cite{SumpterNoidNeuralNetworks1992,Neuralnetworks_BehlerParrinello2007,bpkc2010}. Recently, the focus has been shifting towards emphasizing data efficiency of QML force fields. For example, the symmetrized gradient-domain machine learning (sGDML) model~\cite{chmiela2019sgdml} is able to reconstruct global force fields of small molecules at the ``gold standard'' coupled cluster level of quantum chemistry (demonstrated so far for molecules with up to 25 atoms). The sGDML model achieves a typical accuracy of 0.2 kcal/mol in energies and 1 kcal/mol/{\AA} in atomic forces when using only a few hundreds of molecular conformations obtained from an MD trajectory. Such an accuracy turns out to be crucial when modeling conformational transitions and vibrational spectroscopy of even rather small molecules such as ethanol and aspirin \cite{sauceda2019molecular}. Hence, in the domain of dynamics of small molecules, QML has already enabled essentially exact MD simulations corresponding to a full QM treatment of both electrons and nuclei. 
 Likewise, MD simulations of materials have largely benefited from the application of QML approaches. A recent highlight is the QML-enabled simulation of the growth mechanism of tetrahedral amorphous carbon~\cite{deringer2018computational,caro2018reactivity}, which demonstrates the potential to develop a transferable DFT-level force field for studying complex processes in elemental solids.    
 
 In light of such substantial theoretical advances it is reasonable to wonder whether QML based predictions compare favorably to experimental observables, and if they could even assist with the analysis and possibly even help in the decision making process of the design of new experiments. In this context it is encouraging to note that QML based MD is already beginning to reach the accuracy and efficiency necessary to predict experimental outcomes. For example, Ref.~\cite{chmiela2018towards} uses path-integral molecular dynamics using the sGDML QML force field fitted with CCSD(T) atomic forces to demonstrate that low-frequency excitations of ethanol arise from highly anharmonic combination of vibrational normal modes, thus resolving a long-standing experimental controversy (See Fig.~\ref{fig:Applications} and Ref.~\cite{chmiela2018towards} for a detailed analysis).

%SCALABILITY
Scalability of QML models has also been demonstrated. Atom-by-atom based training on small fragments represented by aforementioned AMONS 
was shown to yield promising results for energies, forces, NMR shifts, and other QM properties for systems of increasing size~\cite{Amons}. 
Extensible QML models have also been proposed in Refs.~\cite{DTNN2017,ANI_IsayevRoitberg2017,collins2018constant,unke2018reactive,chen2018atomic}. 
%MULTIPLE LEVELS
Another way to reduce complexity and increase accuracy of QML models consists of combining
various levels of theory for training and testing~\cite{DeltaPaper2015,pilania2017multi,zaspel2018boosting,batra2019multifidelity}. 
For example, it is possible use a lower level efficient electronic-structure method as a baseline to learn QM properties of a much higher
and more expensive level of theory~\cite{DeltaPaper2015}.

%FORCES
Finally, we note that creating universal QML based force-fields trained and applicable across CCS still remains an open challenge due the substantially more diverse range of chemistries and non-local quantum interactions encountered when navigating configurational {\em and} compositional space at once~\cite{MLatoms_2015,RampiMLQMMM,DTNN2017,ANI_IsayevRoitberg2017,SchNet,christensen2018operators,unke2018reactive,jacobsen2018fly}.

\section{QML based insights into CCS}

In the previous section, we have discussed how QML approaches have aided and extended the reach of QM calculations in several important directions. However, the largest potential appeal of QML is to enable new insights into QM properties of molecules and materials, and ultimately enable expedient exploration of CCS and rational design of molecules and materials with tailored properties.
Most commonly ML methods are employed in the sciences and industry for creating highly predictive models. It has been only very recently that ML models were used as a source of inspiration to learn and obtain insights about the unknown underlying inner regularities hidden in data (e.g.~\cite{DTNN2017}). Given that the ML models achieve this task based on rigorous statistical theory~\cite{lapuschkin2019unmasking}, they have become helpful to generate novel research insights (e.g.~\cite{DTNN2017}) or actionable hypotheses (e.g.~\cite{binder2018towards}) that can be exploited in subsequent steps of (experimental) validation and testing. In other words ML modeling has become a powerful and indispensable part of the scientific discovery process itself. 

\begin{figure*}
\centering
\includegraphics[width=15cm]{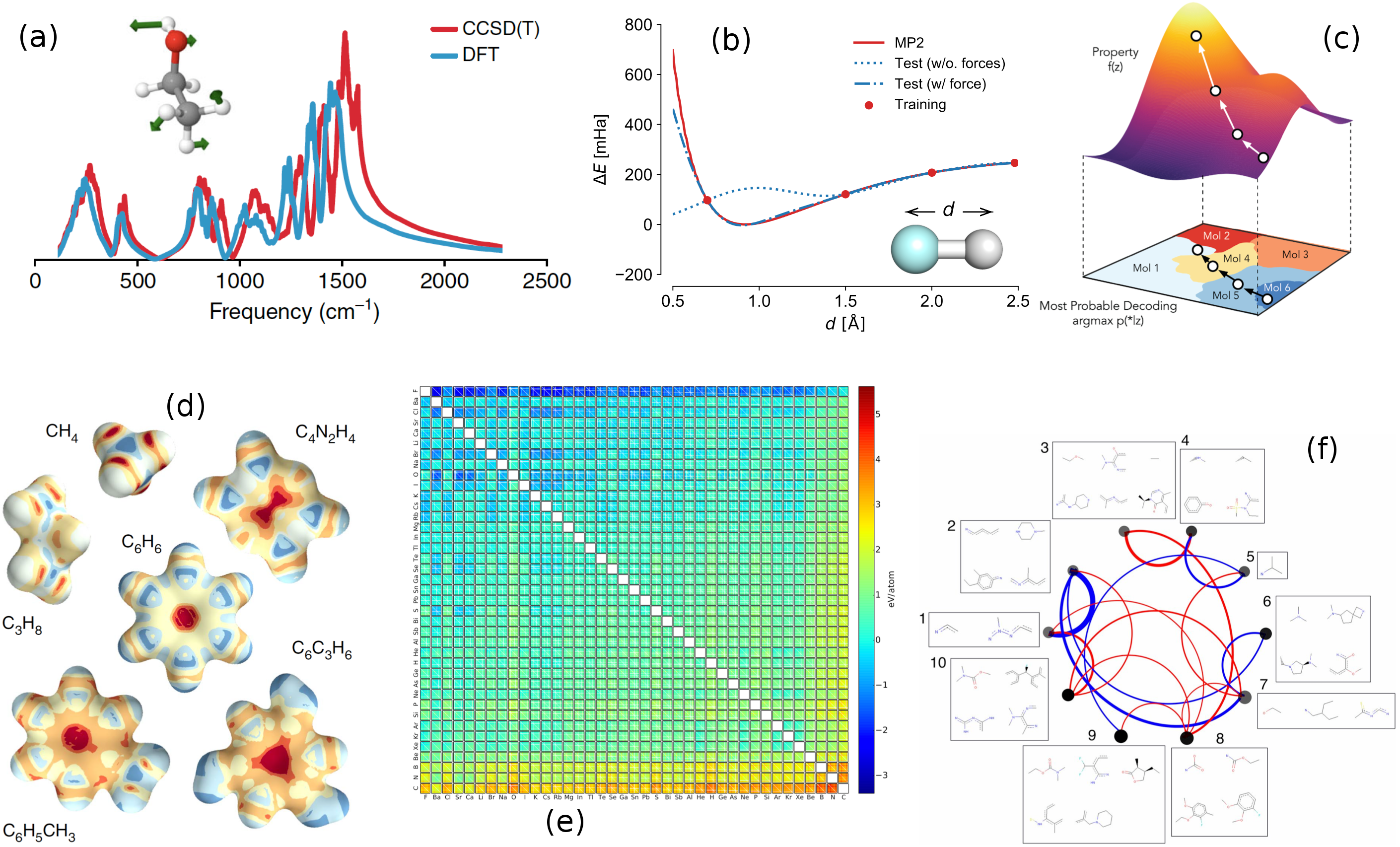}
\caption{
Illustration of insights gained from QML models: 
(a) Dynamical modeling of vibrational anharmonicities in ethanol at CCSD(T) level (picture taken from Ref.~\cite{chmiela2017machine}),
(b) Binding potential of hydrogen fluoride using QML trained on MP2 training data with and without forces (picture from Ref.~\cite{christensen2018operators}),
(c) Smooth property optimization in latent chemical space~\cite{gomez2018},
(d) Spatially resolved chemical potentials inferred for some molecules (picture taken from Ref.~\cite{DTNN2017},
(e) QML based formation energy estimates (color) of $\sim$ 2 million Elpasolite (ABC$_2$D$_6$) crystals made up of main-group elements~\cite{Elpasolite_2016}. 
(f) Identification of (ir)relevant structural motifs for ligand binding~\cite{alphalee2019pnas}.
}
\label{fig:Applications}
\end{figure*}

Fig.~\ref{fig:Applications} illustrates six select chemical examples of how such insights were gained through the help of QML, coming partially from the authors' work, as well as from other literature.
First, the development of an accurate and data efficient sGDML force field model~\cite{chmiela2017machine,chmiela2018towards} (see Fig.~\ref{fig:Applications}a) has allowed to carry out quantum molecular dynamics simulations with essentially exact atomic forces computed with CCSD(T) level of theory. For the first time, this enabled the computation of very accurate thermodynamic and spectroscopic observables for molecules as large as aspirin without compromising between accuracy of atomic forces and accessible time scale of MD simulations. Future work on scaling the molecular size with frameworks such as sGDML will enable fully predictive MD simulations where both electrons and nuclei are treated rigorously via exact quantum-mechanical equations without introducing unnecessary compromises between accuracy and efficiency of molecular simulations.  

Beyond energies and forces, a general QML model needs to be able to predict accurate electronic response properties (akin to evaluating properties as expectation values of the quantum-mechanical wavefunctions). For example, the application of response operator theory to QML models is illustrated in Fig.~\ref{fig:Applications}b: The QML model of the binding curve of hydrogen fluoride improves qualitatively through the explicit inclusion of its derivative with respect to interatomic distance. The same formalism also improves dramatically the prediction of the electrostatic dipole moment, the electric field derivative of the energy, upon inclusion of the corresponding derivatives~\cite{christensen2018operators}. The ultimate goal of QML should be creating a unified surrogate quantum model that can simultaneously learn many quantum-mechanical properties in a data-efficient and accurate manner.    

In other work by G\'omez-Bombarelli et al.~\cite{gomez2018}, the generation of optimal new compounds was optimized through use of generative models in latent space (Fig.~\ref{fig:Applications}c). 
This approach has resulted in promising light-harvesting materials candidates. 
Conceptually, this way of tackling the ``inverse problem'' (property $\rightarrow$ compound)~\cite{zunger2018inverse} through an intermediate step of effectively coarse-graining chemical space is intriguing. 
It constitutes an alternative to the more conventional approaches which 
iteratively solve the ``forward problem'' (compound $\rightarrow$ property)
using gradient based~\cite{Beratan1996,anatole-prl2005,RCD_Yang2006}, Monte Carlo~\cite{ZungerNature1999}, or genetic algorithms~\cite{NorskovPRL2002,zunger2018inverse} or combinations thereof~\cite{AvezacZunger-prb2008}. 
Furthermore, the problem of generating meaningful compounds is solved elegantly in an implicit fashion by directly training on valid SMILES strings, automatically ensuring that the grammar of newly generated SMILES is not being violated. 

Another way to obtain new insights is by analyzing what the QML models have learned from the data, in the spirit of explainable artificial intelligence (AI) where machine learning models are dissected to analyze their inner mechanisms that lead to their respective predictions (see e.g. \cite{bach2015pixel,ribeiro2016should,montavon2018methods,lapuschkin2019unmasking} and references therein). A suitable example of this concept is the analysis of the molecular representation learned by deep tensor neural networks (DTNN)~\cite{DTNN2017}. DTNN models ({\color{black} and other flexible nonparametric ML models}) are trained on QM molecular energies and, in the limit of infinite data, would be able to learn the exact mapping between molecular structures and the solution of the Schr\"odinger equation. Since the exact solution can only be formally achieved in the limit by exactly representing the wavefunction, there is mathematically no other choice for DTNN than forming an exact representation of the wavefunction. In practice, the representation is trained on a finite number of molecules, hence DTNN learns the ``Schr\"odinger  mapping'' on a finite set of molecules which does not necessarily need to be a one-to-one representation of the wavefunction. One can query the learned representation by adding a probe atom to a given molecule~\cite{DTNN2017}. By visualizing the energy isosurface of the probe atom, one can immediately see that the obtained representation exhibits features that closely resemble electron densities or electrostatic potentials, indicating that the model is able to infer QM features in the representation directly from a restricted set of QM energies (see Figure ~\ref{fig:Applications}d). Hence, the DTNN approach is attempting to solve an inverse problem \cite{franceschetti1999inverse} of constructing a coarse-grained QM representation from a finite set of molecular energies or other QM properties (see Fig.~\ref{fig:Applications}d). 
Despite being trained only on total energies of molecules, the DTNN approach clearly grasps fundamental chemical concepts such as bond saturation and different degrees of aromaticity. For example, the DTNN model predicts the C$_6$O$_3$H$_6$ molecule to be ``more aromatic'' than benzene or toluene~\cite{DTNN2017}. It turns out that C$_6$O$_3$H$_6$ does have higher ring stability than both benzene and toluene and DTNN predicts it to be the molecule with the most stable aromatic carbon ring among all molecules in the QM9 database~\cite{DTNN2017}. Interestingly, the mathematical construction of the DTNN model ({\color{black} and other flexible nonparametric models based on atomic contributions}) provide statistically rigorous partitioning of extensive molecular properties into atomic contributions -- a long-standing challenge for quantum-mechanical calculations of molecules. 

{\color{black} We note that while DTNN is in principle able to construct a ``Schr\"odinger  map'' between molecular Hamiltonians and molecular quantum properties, detailed analysis of the underlying representation learned by DTNN amounts to a complex inverse problem. To address this problem, some of us have recently developed a generalized SchNOrb architecture that learns the DFT wavefunction directly~\cite{schnorb}. As an outlook into future work, it would be desirable to unify both ML architectures (DTNN/SchNet and SchNOrb) to gain enhanced understanding into quantum mechanics of molecules by combining the direct ``Schr\"odinger  map'' (SchNOrb) with the inverse ``Schr\"odinger  map'' (DTNN/SchNet).}

The QML approach is evidently also applicable to solids. In Ref.~\cite{Elpasolite_2016}, formation energies of $\sim$2 million Elpasolite crystals (of ABC$_2$D$_6$ sum formula) made up of main group elements were estimated (see Fig.~\ref{fig:Applications}e). Ranking by estimated thermodynamic stability on the convex hull resulted in the identification of nearly one hundred novel crystals, all expected to be stable, and subsequently added to the Materials Project data base~\cite{MaterialsProject}. Furthermore, detailed analysis of oxidation states resulted in the discovery of an exotic crystal, NFAl$_2$Ca$_6$, in which Al carries an unusual negative oxidation state -- this surprising find was made possible only through the systematic combination of quantum-mechanical calculations and machine learning. 

The fundamental nature of QML is not restricted to pre-calculated datasets. 
Within seminal work, Lee and others have applied ML to experimental data in order to understand and enhance ligand-protein binding~\cite{alphalee2019pnas}, shown in Fig.~\ref{fig:Applications}f.
In particular, random matrix theory was used to identify those chemical groups and features which do and do not strongly affect binding. Such analysis can provide invaluable information on how to exploit local chemistries in order to steer a complex chemical property such as drug-target binding.  

All of these examples demonstrate the great potential of QML for extracting statistical insights and new knowledge about quantum property relationships throughout CCS, which are not directly available through conventional quantum calculations.

\begin{figure}
\centering
\includegraphics[width=4cm]{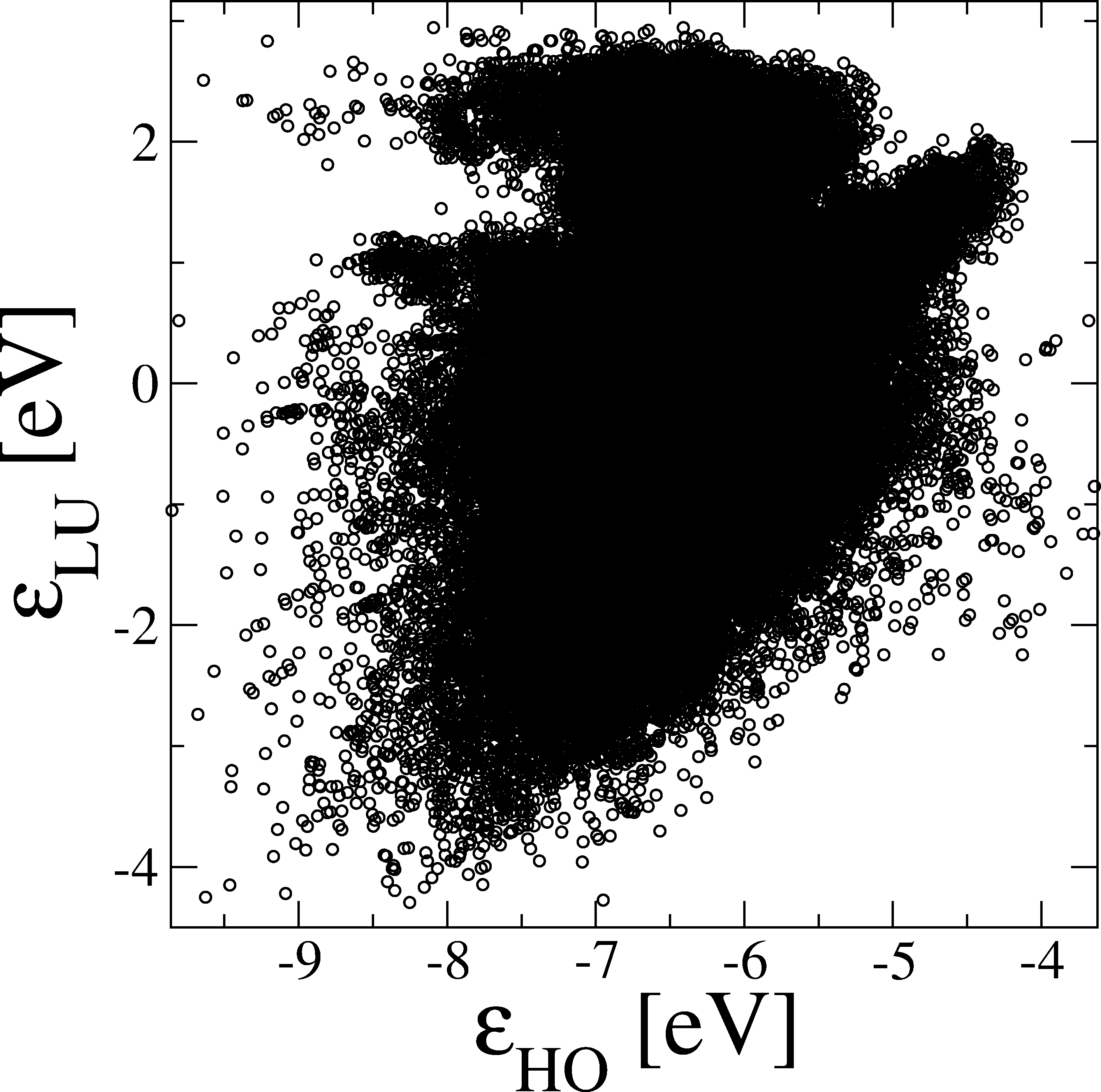}
\includegraphics[width=4cm]{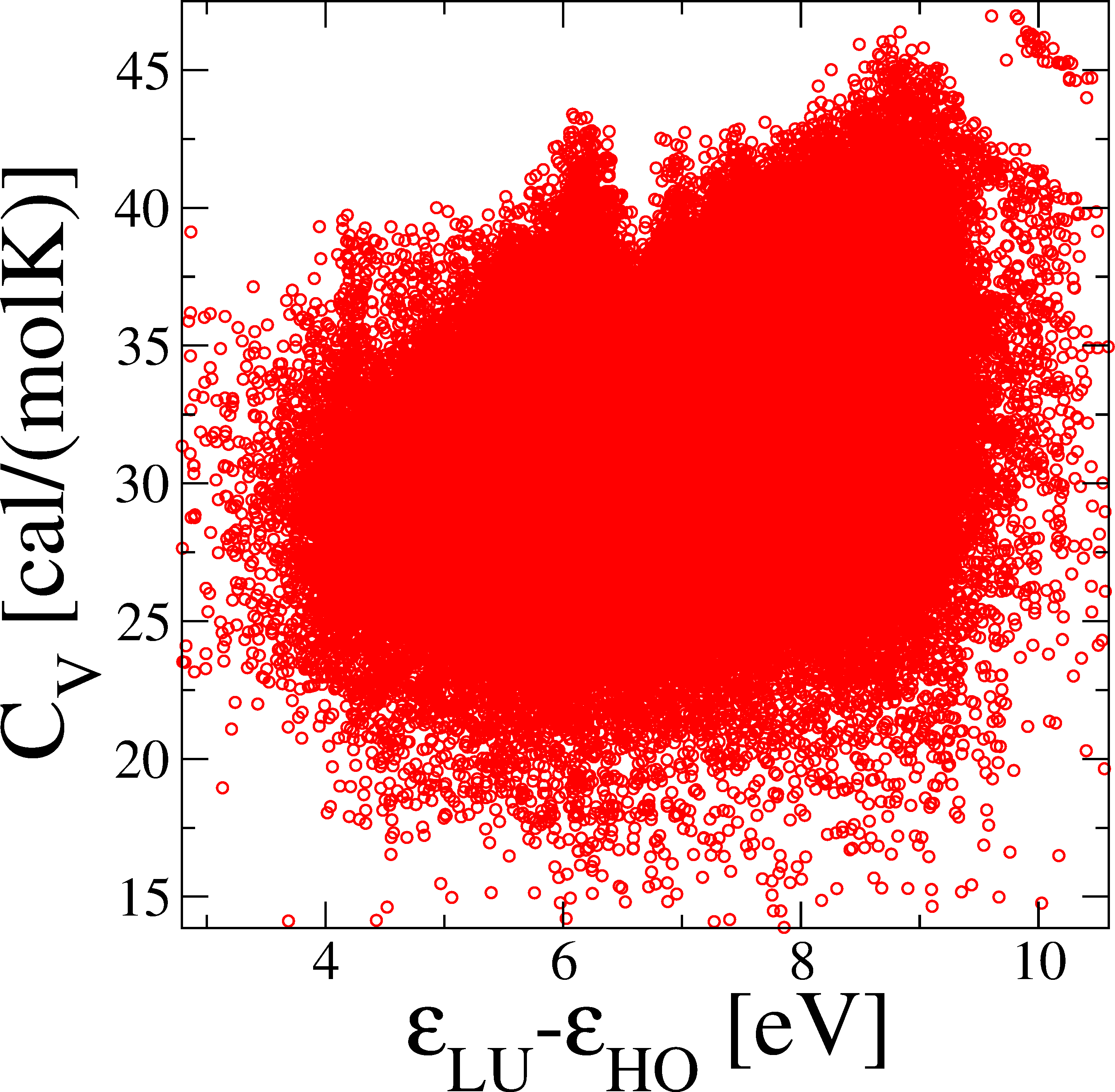}
\includegraphics[width=4.2cm]{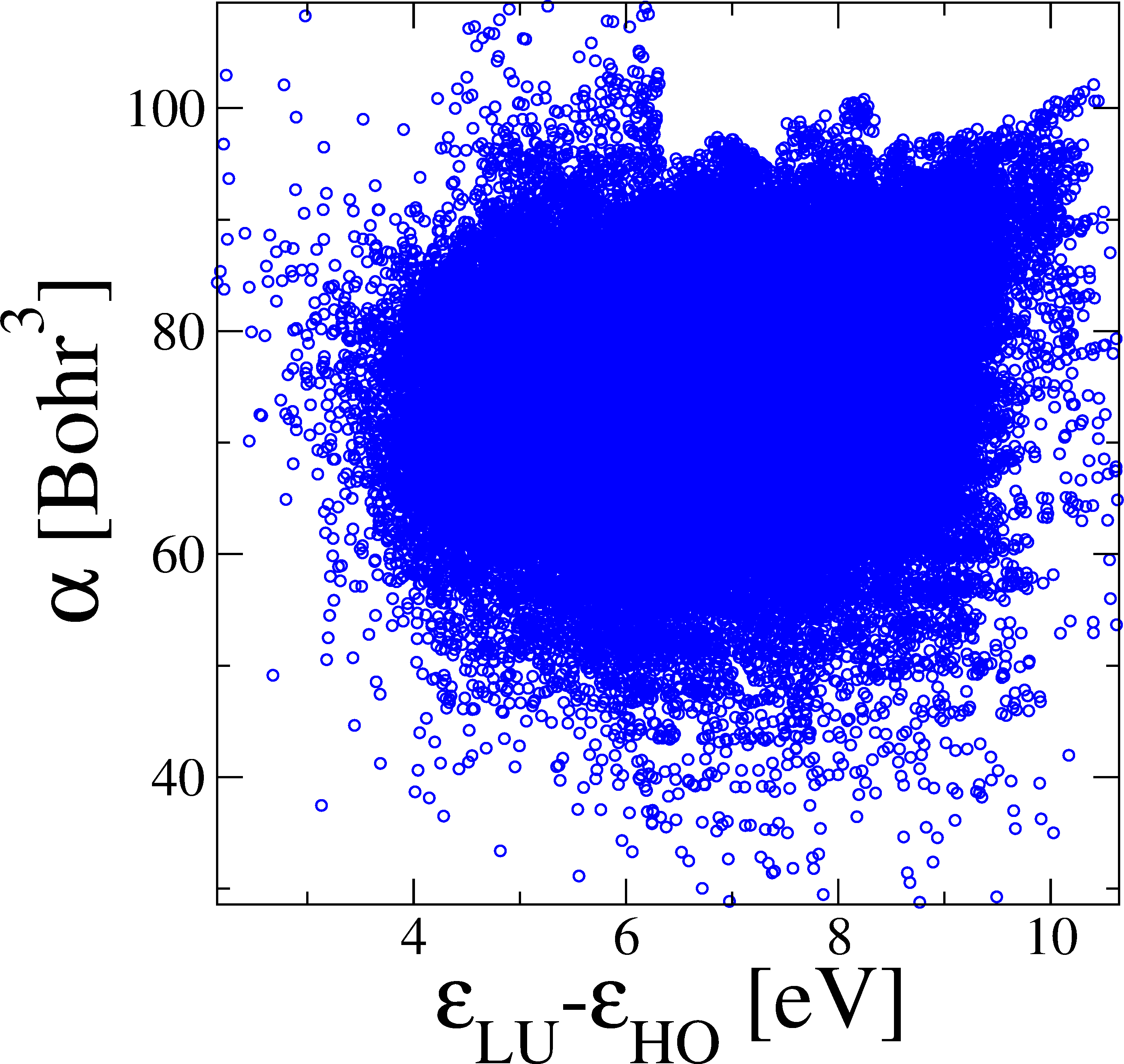}
\includegraphics[width=4.2cm]{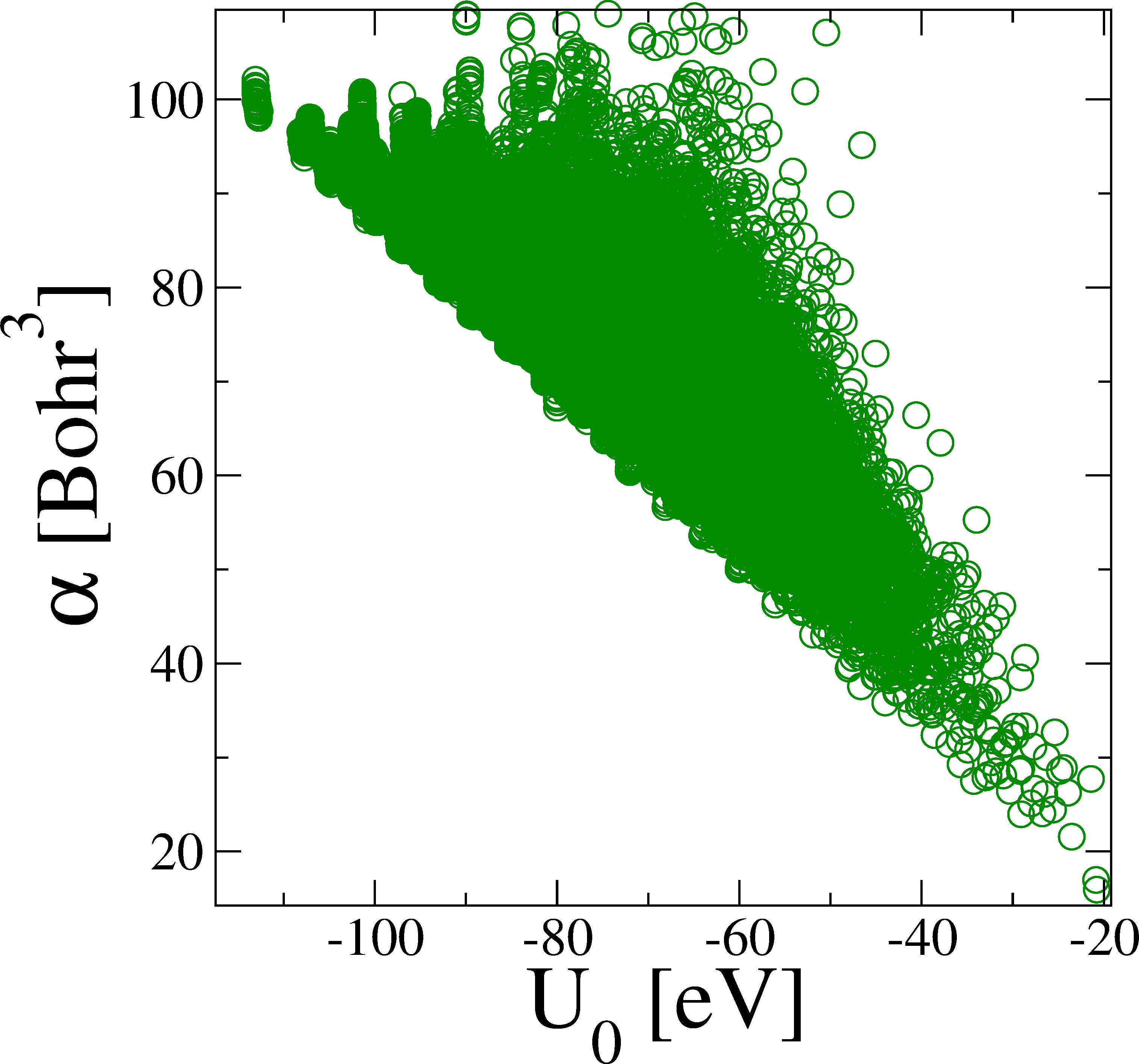}
\caption{
Weak dependency of property pairs in CCS illustrated by 
eigenvalues of highest occupied (HO) and lowest unoccupied (LU) molecular orbitals (top left), 
heat-capacity and electronic gap (top right),
polarizability and electronic gap (bottom left), 
polarizability and atomization energy (bottom right).
See also Ref.~\cite{Montavon2013} for further examples.}
\label{fig:Freedom}
\end{figure}

The data-driven nature of QML approach based on exploring increasingly larger swaths of CCS also enables new insights into the possibility of rational design of molecules with multiple desired properties. For example, in a hypothetic drug design scenario, one could be interested in finding a particularly stable molecule with a large polarizability $\alpha$ (hence stabilizing drug-protein van der Waals interaction) and a large electronic HOMO-LUMO gap $E_{\rm{gap}}$ (exhibiting stability to external electrostatic fields). These three requirements would normally be considered contradictory to each other. Firstly, stability is typically inversely correlated with polarizability -- stable molecules are normally thought to have small polarizability~\cite{hohm1994dipole,hohm2000there,Geerlings_DFTConcepts}. 
Secondly, HOMO-LUMO gap is the leading-order contribution that appears in the denominator of the polarizability formula, hence it is often assumed that polarizable molecules should have small HOMO-LUMO gaps. One is then faced with a difficult question of whether the formulated design problem of low $E$, high $\alpha$, and high $E_{\rm{gap}}$ is hopeless? This question can be partially answered by analyzing the pairwise correlation between different molecular properties for a large but finite set of drug-like molecules. These correlations are shown in Figure~\ref{fig:Freedom} for roughly 7k molecules in the QM7 dataset. The first observation is that the correlation between most electronic properties is rather weak, if at all present. Most strikingly, above the lower bound of polarizability and atomization energy (both must be bounded from below) we observe no visible correlation. The same observation is made for polarizability vs. HOMO-LUMO gap. This pairwise comparison of three different properties leads us to the suggestion that one can find many drug-like molecules that satisfy the seemingly contrasting requirements of high stability, high polarizability, and a large HOMO-LUMO gap. Similar ``freedom of design'' is observed for HOMO vs. LUMO eigenvalues, as well as for HOMO-LUMO gap vs. heat capacity.
This data-driven analysis, spurred by the QML approach, illustrates a novel way to look at rational design in CCS, breaking conventional descriptor-property rules as well as notions of restricted chemical diversity.

\begin{figure}
\centering
\includegraphics[width=8.5cm]{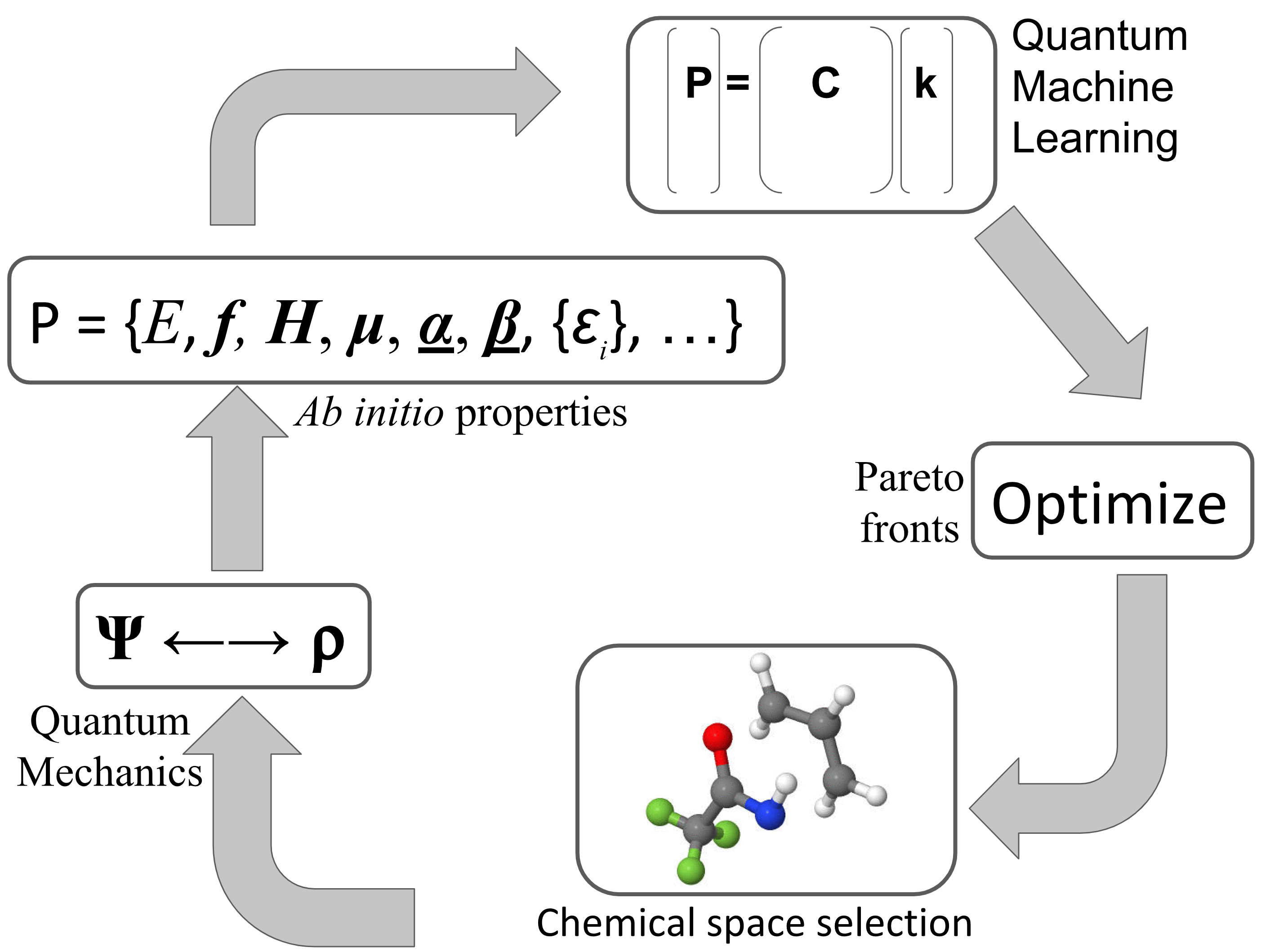}
\caption{Application concept of QML:
Multiple properties can be iteratively optimized in chemical space, spanned by
structural and compositional degrees of freedom.  
After selection of some initial representative target compound(s), 
quantum mechanics is invoked to calculate relevant {\em ab initio} properties 
and to subsequently generate (update) a QML model. Multi-objective property
optimization algorithms can subsequently create a new list of candidate structures until convergence.
}
\label{fig:Design}
\end{figure}

These developments result in a paradigm shift in atomistic simulations, due to (a) usage of rigorous QM/SM priors and data, instead of heuristic cheminformatics, (b) a holistic explorative and comprehensive view on CCS, rather than the traditionally assumed encyclopedic view where each process is studied one compound at a time, and (c) insights provided by novel tools relevant for several scientific communities, i.e. electronic structure prediction, material science, organic chemistry, molecular dynamics and drug discovery. 
Figure~\ref{fig:Design} illustrates how this new view could be leveraged for boosting more conventional computational compound design applications, and contribute substantially to on-going experimental efforts, as also recently reviewed in the context of catalyst design~\cite{batista2019designMLalchemy}.
As such, QML is clearly already taking the first steps in the direction of generally tackling the inverse
design question in CCS~\cite{zunger2018inverse,franceschetti1999inverse,jorgensen2018exploration}.

Furthermore, we are confident that QML is not just limited to direct and inverse design, but it amounts to a fresh view on molecules and materials, and it opens many possibilities to be explored. For example, QML represents a unique opportunity to rigorously test and assess known rules in chemistry, derived from human intuition and empiricism, up to an unprecedented degree of statistical confidence if sufficient data is made available. Furthermore, QML models could also help us discover and extract new concepts, which have hitherto escaped the notion of chemists. These developments may provide a further boost to attempts to gain a holistic understanding of CCS, its structure, and what it holds in store for us in terms of interesting materials, properties, or processes.

\section{Challenges and Outlook }

While the QML field has seen a tremendous progress over the past few years, many more challenges remain to be addressed. In the following, we proceed by charting some of the challenges we consider as most interesting and pressing.

\subsection{Towards Big Data in CCS}
An important limitation that the QML field is slowly encountering is the lack of large comprehensive datasets. While datasets like QM7 \cite{CM,Montavon2013}, QM9 \cite{QM9}, materials project \cite{MaterialsProject}, OQMD~\cite{OQMD}, Elpasolites~\cite{Elpasolite_2016}, MD17~\cite{chmiela2017machine},  ANI-1 dataset~\cite{smith2017ani}, Silicon structures~\cite{bartok2018machine},
%{\bf ALEX could you please ADD MORE References in the paragraph} 
have served well for the development and testing of novel ML techniques, there is an inherent danger to overfit to benchmarks -- a fact that has driven e.g. the field of computer vision to exploring larger and ever more complex scenarios (e.g. \cite{deng2009imagenet,rohrbach2012database}). It will be therefore important to establish high quality large scale data resources that can enable a new model generation that increasingly explores both compositional and configurational aspects of CCS~\cite{DTNN2017,ANI_IsayevRoitberg2017}.

However, while having comprehensive big data in CCS would be helpful, due to the combinatorially scaling CCS it remains crucial to develop efficient models that only rely on small amounts of data. Clearly, learning with abundant data is straightforward, but it becomes more challenging to reliably learn from small datasets. Here it becomes crucial to include prior physics-based knowledge and invariance information to become more data efficient without compromising the robustness and accuracy of the QML model~\cite{chmiela2019sgdml,chmiela2018towards,chmiela2017machine}. 

In addition, this new generation of models should ideally quantify the uncertainty of its own prediction~\cite{ML4Kieron2012, schwaighofer2009wrong,smith2013uncertainty} possibly in combination with active learning strategies that may lead to improved sampling of CCS and effectively lead to lower model uncertainties~\cite{smith2018less,gubaev2018machine}. It is important to note that active learning actually induces non-stationarity in learning \cite{sugiyama2012machine}. In addition, models will need to be able to explain their prediction (e.g.~\cite{bach2015pixel,lapuschkin2019unmasking}) such that a  model will not only serve as predictor but in addition as a source of generating insight \cite{DTNN2017,SchNet}. 
In other words, novel developments need to consider data generation, model building, explanation, insight extraction and sampling in a single comprehensive framework. 
The recently introduced AMON approach (see Fig.~\ref{fig:Constellation}), which selects molecular fragments and trains QML models on the fly, represents a first step in this direction~\cite{Amons}.

\subsection{Learning Complex Electronic Properties}
Furthermore, current limitations and shortcomings of QML models include predicting intensive properties such as the eigenvalues of molecular orbitals~\cite{HDAD}, or excited state properties such as excitation energies~\cite{ML-TDDFTEnrico2015,pronobis2018capturing}. 
Transferable yet accurate QML models of electron densities~\cite{grisafi2018transferable} and
molecular orbitals in molecules and band structures in solids also remain a challenge.
Another issue, lurking behind rigorous and robust statistical learning procedures, such as $k$-fold cross-validation and converged
learning curves, is the selection bias encoded in many of the
training sets employed in the field: Stability or property distributions are typically unknown in CCS, and therefore hamper the rigorous assessment of the degree to which any given data set is truly representative of broader chemical spaces. 
Similar problems of representability were also encountered when trying to measure how accessible website content is reflected in search results when using different search engines \cite{lawrence1999accessibility,lawrence1998searching}.

Other challenges include (i) the determination of the irreducible set of variables (formal scaling is only an upper bound but what is effective dimensionality? (see e.g.~\cite{Braun2008}), (ii) the question of how to render prediction errors constant throughout CCS, and (iii) a quantitative understanding of the relation between the QML models' learning efficiency, as manifested in learning curves, and the dimensionalities of CCS as encoded in the training data.

\subsection{Multiscale QML models}
A very promising research direction is the integration of QML models across different levels of theory.
Exploiting decades of research on the validity and applicability of the various approximations made when solving the Schr\"odinger's equation, ample data obtained with computationally less demanding approximations can be combined with fewer but more accurate data points. As a result, the QML models must only learn the differences between the various levels of theory, which is substantially less demanding in terms of data needs. As such, these $\Delta$-learning approaches allow to invest the model complexity on the truly difficult aspects~\cite{ginzburg1994combined,DeltaPaper2015,pilania2017multi,zaspel2018boosting,Bogojeski2019,IsayevNatComm2019}. 

Many studies so far have successfully explored structure-property relationships in restricted chemical spaces. However, the final goal is to enable global and universal exploration of CCS exploiting the appropriate framework offered by the combination of QM, SM, and ML. 
Here, we have connected some of the ongoing efforts in this endeavour and have provided pointers into the broad activities of the scientific field that has emerged. We would like to stress that the tools available now have reached a level of maturity that should become helpful to a wider community of researchers and practitioners.

\subsection{Towards molecular design with QML}
In the following, we would like to outline three concrete open challenges where QML tools still have to be developed in order to find broad application. 
%Among open challenges we can mention extensions of presented QML techniques to statistical mechanical averages and mixtures and larger systems, as well as multi-scale and coarse-grained approximations within CCS, and QML approaches that go beyond the BO-approximation, and successfully
%account for relativistic, nuclear quantum effects, and excited states. 
\begin{itemize}
\item[(i)] {\em More observables} The use of QML in order to estimate statistical mechanics observables, e.g. leading to the prediction free energy profiles of rare events, remains an outstanding challenge. This will enable direct validation by comparison to experimental gas-phase rate constants from the literature, as well as to vibrational or linear free energy perturbation based free energy estimates. In order to compare to liquid chemistry results, one has to also include solvent effects (calculated through continuum solvent models, through addition of shells of solvent, or through periodic boundary conditions). 
If necessary, one should also include nuclear quantum effects through use of path-integral simulations.
These developments will serve the goal of establishing once and for all the validity of the QML based approach by direct comparison to experiment, rather than to pre-calculated quantum results. 
%, from literature, or within collaborative efforts. 
\item[(ii)] {\em Experimental design} 
QML forms the natural basis for software capable to run assisted (automatized/robot) experiments or helping scientists with experimental design decisions. 
Some work along those lines, suggesting new materials, has already been published~\cite{Elpasolite_2016,NorskovML2016,gomez2018,clemence2018catalystML}. 
Latent space applications and computational alchemy can be combined with state-of-the-art optimizers, in essence paving the way towards the experimental realization of aforementioned multi-property design tasks relevant to the identification of promising drug, photovoltaic, battery or catalyst candidates.
\item[(iii)] {\em Reaction design using QML.}
Computer based reaction planning and discovery has a longstanding history in chemistry dating back to the sixties, and including contributions by Corey et al in 1972~\cite{CoreySynthesis1972}, or Herges and Hoock in 1992~\cite{HERGES1992}.
A comprehensive review of the field has recently been published~\cite{Chematica2016},
also discussing the {\tt Chematica} software 
which proposes new synthetic routes based on new combinations of already established reactions reported in the literature.
Laino and co-workers~\cite{schwaller2018found} and Segler et al~\cite{segler2018planning} also
introduced literature based ML models for chemical reactions.
Such approaches can be problematic, however, when it comes to combinations or reaction conditions
for which previously published reactions are not representative. Moreover, the approach is inherently biased
to historically known chemistries and combinations thereof, excluding the possibility to discover entirely new reaction mechanisms and synthesis pathways.
The latter point, however, is critical for example in the context of developing new catalysts, and more fundamentally
to fill any existing "gaps" of our understanding of potentially useful reactions.
In order to computationally predict novel reaction profiles we must rely on universal first principles
based numerical simulation of the relevant quantum and statistical mechanics which accounts for
the electronic and atomic movements resulting in the relevant reaction energies and barriers~\cite{Leach,MolecularElectronicStructureTheory,tuckerman_book_SM}.
QM and SM represent the appropriate physics framework necessary to describe the
electronic rearrangements occurring during a reaction {\em and} for the freedom to dial in atomic configurations and chemical composition at will.
When optimizing reactions through screens of prospective combinations of reactants, products, and external conditions, the role of solvents and catalysts is crucial as they can alter the ranking and even render reactions realistic which otherwise would have been impossible. One therefore has to expand training sets to include libraries of solvents and simple catalysts, and  apply extended QML models not only of energies and forces but also of statistical mechanical averages. 
Once trained, these QML models could be used to optimize reaction conditions (solvents, ions, temperature, pressure) in chemical space  using gradient-free optimizers such as Monte Carlo/Genetic/or simplex algorithms. First steps in this direction were already taken in 2012 by Rupp et al.~~\cite{ML4Graeme2012}.
\end{itemize}

\section{Conclusions}

Over recent years, overwhelming evidence has been gathered by the community suggesting that QML models can truly generalize throughout CCS. As such, a paradigm shift has occurred when moving from globally fitting parameters in fixed functional forms, inspired by physics informed models (such as UFF~\cite{UFFRappeGoddard1992} or semi-empirical methods (such as PM6~\cite{PM6,PM7} or DFTB~\cite{DFTB+}), to locally optimizing regression weights in generic basis set expansions which can be converged in size. Resulting QML models enable (a) rapid predictions of relevant quantum properties for new out-of-sample systems (after training) and (b) converged predictive power through sufficiently large training sets (as evinced by convergence properties of learning curves). 
Thanks to the tremendous reduction in computational cost of query tasks, QML models enable to shift the focus away from studying individual instances in CCS towards entire ensembles of compounds. 
Recovery (or rejection) of known, and discovery and elucidation of unknown structure-property relationships has therefore become a feasible, realistic, and valuable goal which was previously not accessible. 

Conceptual challenges include the definition of locality in CCS, i.e.~when the QML models are interpolating or rather when their validity fades. While ensemble methods and Gaussian Processes provide a first direction of uncertainty quantification in a limited domain of applicability, mathematically and physically more well founded methods are still waiting to be discovered.  
Rigorous definitions of diversity (e.g.~\cite{marienwald2018tight}), properly rooted in QM and SM, might also be necessary to tackle selection bias problem and to maximize data efficiency. 
First principles based diversity measures would have to properly account for all sorts of systems including metal organic frameworks (MOFs), nano-materials, organic materials, functional materials, inorganic crystals, metastable solids, liquid mixtures, or bio-systems.

An educational challenge corresponds to the establishment of the academic curriculum for this interdisciplinary young field, where chemistry, physics, and computer science have to develop tightly interwoven research programs. Conventional curriculae in traditional departments of chemistry, materials science, physics, computer science, or biology departments do not cover the course-work necessary for students to appropriately reach a level where they can meaningfully contribute to this line of research. 
%Chemical diversity vs transferability and universality of models. E.g. DFT is mediocre for many situations, ML is excellent for some and poor for others … how can we turn ML into a model which outperforms DFT throughout? Focus on understanding, rather than blind ML. 
%New synthetic techniques (e.g. atomic layer deposition) open up new regions in CCS … 

Finally, we would like to stress that the progress made and described herein is dwarfed by the scope of the problem: Gaining virtual control of CCS through physics based understanding has remained elusive for all of humanity's past scientific efforts. One of the many rewards of reaching this goal would be the routine discovery and design of interesting molecules and materials with desired properties. 
As such, the community has so far just been scratching the surface of what is to come. To further push the frontier of this field of science, sustained and increasing investments are necessary in terms of computer power, interdisciplinary education and training, funding agencies and most importantly: human interdisciplinary creativity. 

\section*{Acknowledgement}
All authors thank J. Wagner, F. A. Faber, and K. Sch\"utt  for preparing the graphics in Figs.~1,2, and 3, respectively.
O.A.v.L. acknowledges funding from the Swiss National Science foundation (No.~PP00P2\_138932 and 407540\_167186 NFP 75 Big Data) and from the European Research Council (ERC-CoG grant QML). This work was partly supported by the NCCR MARVEL, funded by the Swiss National Science Foundation. A.T. acknowledges financial support from the European Research Council (ERC-CoG grant BeStMo). K.-R.M. acknowledges partial financial support by the German
Ministry for Education and Research (BMBF) under
Grants 01IS14013A-E, 01GQ1115 and 01GQ0850;
Deutsche Forschungsgesellschaft (DFG) under
Grant  Math+, EXC 2046/1, Project ID 390685689 and by the Technology Promotion (IITP) grant funded by the
Korea government (No. 2017-0-00451, No. 2017-0-01779). Correspondence to OAvL, KRM and AT.

\appendix
\section{Appendix: Software}
Here  some pointers are provided to software packages of the discussed machine learning methods for quantum chemistry. General code for running QML is found in \cite{qmlcode2017}. Deep learning methods such as DTNN and SchNet can be readily implemented using the SchNetpack~\cite{schnetpack}. The technical intricacies of sGDML where prior information has been included in the learning system (i.e.~symmetries, sampling and energy conservation) is readily and easily usable in~\cite{chmiela2019sgdml}. The iNNvestigate toolbox that allows to explain nonlinear learning methods such as deep learning is described in~\cite{alber2019innvestigate}.

\bibliography{literatur}{}
\bibliographystyle{ieeetr}

\end{document}